\documentstyle[aps,epsfig]{revtex}
\draft
\def\la{{\langle}}
\def\ra{{\rangle}}
\def\vep{{\varepsilon}}
\newcommand{\beq}{\begin{equation}}
\newcommand{\eeq}{\end{equation}}
\newcommand{\beqa}{\begin{eqnarray}}
\newcommand{\eeqa}{\end{eqnarray}}
\newcommand{\da}{^\dagger}
\newcommand{\wh}{\widehat}
\newcommand{\Op}{{\wh{\Omega}_+}}
\newcommand{\Om}{{\wh{\Omega}_-}}
\newcommand{\Opm}{{\wh{\Omega}_{\pm}}}

\newcommand{\intf}{\int_{-\infty}^\infty}

\begin{document}

\title{Generalizations of Kijowski's time-of-arrival 
distribution for interaction potentials}
\author{A. D. Baute$^{1,2}$, I. L. Egusquiza$^1$, and J. G. Muga$^{2}$}
\address{$^1$ Fisika Teorikoaren Saila,
Euskal Herriko Unibertsitatea,
644 P.K., 48080 Bilbao, Spain}
\address{$^2$ Departamento de Qu\'\i mica-F\'\i sica,
Universidad del Pa\'\i s
Vasco, Apdo. 644, 48080 Bilbao, Spain}
\date{\today}
\maketitle
\begin{abstract}
Several proposals for a time-of-arrival distribution of
ensembles of independent quantum
particles subject to an external interaction potential
are compared making use of the ``crossing state'' concept.
It is shown that only one of them has the properties 
expected for a classical distribution in the classical limit.
The comparison is illustrated numerically with a collision 
of a Gaussian wave packet with an opaque  
square barrier.   
\end{abstract}
\pacs{PACS: 03.65.-w\hfill EHU-FT/0101}

\section{Introduction}
In many experiments, the observed quantities are the instants 
of occurrence of certain events, or the durations of processes.
However, the standard quantum formalism does not provide, at least in an   
obvious manner, a working rule for time observables. 
In spite of the difficulties and objections to consider
time as a quantum observable (by Pauli \cite{Pauli58},
Allcock \cite{Allcock69},
and other authors),
many researchers have sporadically tried to fill this theoretical 
lacuna. The effort in that direction has become more 
intense and systematic in recent years. 
In particular, much attention has been devoted to the quantum description 
of the ``arrival time''
\cite{Kijowski74,Werner86,MBM95,AOPRU98,Leavens98,Delgado98,MSP98,MLP98,%
Finkelstein99,Toller99,Galapon99,Halliwell99ba,Kijowski99,ORU99,Delgado99,%
MPL99,KW99,LJPU99,LJPU00,Leon00,BSPME00,ML00,BEMS00,EM00a,EM00b}, 
see \cite{ML00} for a recent review.  In its simplest form, the
problem is to define an ideal quantum arrival time distribution for a
wave packet moving freely in one dimension.  By imposing a number of
classically motivated conditions (normalization, positivity, minimum
variance, and a certain symmetry with respect to the arrival point
$X$) this is solved uniquely by Kijowski's distribution
\cite{Kijowski74}.  This distribution may also be associated with the
positive operator valued measure (POVM) generated by the improper
eigenstates of the time-of-arrival maximally symmetric operator of
Aharonov and Bohm \cite{AB61}, $\wh{T}_{AB}$,
\beq
\wh{T}_{AB}=\frac{m X}{\wh{p}}-\frac{m}{2}\left(\wh{x}\frac{1}{\wh{p}}
+\frac{1}{\wh{p}}\wh{x}\right), 
\eeq
where $\wh{x}$ and $\wh{p}$ are position and momentum operators and 
$m$ is the mass. 
Aside from the 
discussion of remaining interpretational
puzzles, an important pending question is its generalization for
particles affected by interaction potentials 
\cite{ML00}. However, Kijowski's set of conditions cannot be applied
in the general case \cite{Kijowski74}, where, classically, not all
particles necessarily arrive, or multiple arrivals may have to be
considered.  Neither can  the simple symmetrization rule leading to
$\wh{T}_{AB}$  be used \cite{Galapon00}.  Thus a
generalization of $\Pi_K$ required some novel approach and it is only
very recently that some of them have been explored, following
different heuristic and/or formal arguments.  Their physical content must be
analyzed in order to select one that does fit into the proposed
objective (although several might be adequate).  This is the aim of
the present paper, where we shall examine three possible
generalizations of Kijowski's distribution.  The unifying framework is
provided by the concept of ``crossing state'' introduced in
\cite{BSPME00}, and inspired by Wigner's formalization of the
time-energy uncertainty principle \cite{Wigner72}.  In all the
proposals examined here the (candidate) time-of-arrival distribution
at point $X$ may be obtained from the overlap of the time dependent
wave function and the crossing states $|u^\beta (X)\ra$ as
\beq\label{gen}
\Pi(T)=\sum_\beta \Pi^\beta(T)=\sum_\beta|\la \psi(T)|u^\beta(X)\ra|^2,
\eeq
where $\beta=L,R$ is an index for ``left'' and ``right'';   
its exact meaning will vary in the different 
generalizations.
We shall use a unified notation that will differ in general
from that of the original papers
to facilitate the common presentation and comparison.
Incidentally, all distributions defined in this manner are automatically
covariant with respect to time translations, namely, the arrivals predicted 
for a given fixed instant are independent of the choice made for the 
origin of time \cite{BEMS00,ML00}.

\section{Different crossing states}

\subsection{Free motion}
Kijowski's distribution is obtained from the general expression 
(\ref{gen}) with the following crossing states
(their coordinate representation, time evolution, 
and wave packets peaked around them have  
been studied in \cite{MLP98}),
\beq\label{uK}
|u^\beta(X)\ra_K=(|\wh{p}|/m)^{1/2}\Theta(\alpha \wh{p})|X\ra, 
\eeq
where
\beq
\alpha=\cases{
+ & ${\rm if}\; \beta=L$\cr
- & ${\rm if}\; \beta=R$,\cr}
\eeq
$|\wh{p}|^{1/2}$ is defined by its action on plane waves, 
\beq
|\wh{p}|^{1/2}|p\ra=|p|^{1/2}|p\ra,
\eeq
(the positive root is taken), and $\Theta$ is the Heaviside distribution.
 
Now we can write Kijowski's distribution in operator form as  
\beq\label{Kijo}
\Pi_K(T)=\sum_\alpha
\la \psi(T)|\left[\Theta(\alpha \wh{p}){(|\wh{p}|/m)^{1/2}}\delta(X-\wh{x})
{(|\wh{p}|/m)^{1/2}}\Theta(\alpha \wh{p})\right]|\psi(T)\ra.  
\eeq
Each of the operators in brackets (with $\alpha=+$ or $-$ respectively) 
corresponds classically, i.e., disregarding the lack of commutativity
between $\wh{x}$  and $\wh{p}$, to the classical dynamical variable
\beq\label{jmod}
\delta(x-X)\frac{\alpha p}{m}\Theta(\alpha p),     
\eeq
whose average $\la .... \ra_{cl}$ over a classical phase space density
gives the flux due to particles arriving from the left,
$J_{cl}^L$, 
or minus the flux due to particles arriving from the right, 
$-J_{cl}^R$ (which is a positive quantity, $-J_{cl}^R>0$),
\beq
\alpha J_{cl}^\beta=
\la \delta(x-X)\frac{\alpha p}{m}\Theta(\alpha p)\ra_{cl},
\eeq
in other words, the modulus of the 
flux of particles of the classical ensemble that arrive from one side at 
a given time \cite{Delgado98}.
The addition of these two contributions is the classical time-of-arrival
distribution, 
\beq\label{picl}
\Pi_{cl}(T)=J_{cl}^L-J_{cl}^R
\eeq

\subsection{Interacting case: first proposal}
The previous discussion 
motivates the first generalization of $\Pi_K$ considered here 
for independent particles affected by an arbitrary interaction potential. 
Since (\ref{picl}) is valid regardless of the presence
of a potential, and the dynamical variables for the two 
flux contributions are always given by (\ref{jmod}),    
it was proposed in \cite{BSPME00} that the quantum time-of-arrival
distribution in the general case 
be given by the same expression used in the free motion case,
Eq. (\ref{Kijo}), and by the same crossing states,
$|u^\beta\ra_1=|u^\beta\ra_K$, 
as they lead to operators in correspondence with the
classical expression (\ref{jmod}).
In other words, the proposed distribution is given by 
\beq\label{Pi1}
\Pi_1(T)=\sum_\beta |\la \psi(T)|u^\beta\ra_K|^2,
\eeq
where now $\psi$ evolves with the full Hamiltonian $\wh{H}=
\wh{H}_0+\wh{V}$ which includes a kinetic term, $\wh{H}_0$, and a 
potential term, $\wh{V}$.  We are thus extending to the quantum domain
the fact that in classical mechanics the expressions for dynamical
variables representing the partial fluxes do not vary from the
free-motion case to the interacting case.  Here, the state ($\psi(T)$
quantally, and the evolved phase space density classically), rather
than the dynamical variable, contains the information which is
specific to each particular Hamiltonian.  At the very least, this
generalization of Kijowski's distribution has the merit of being
simple, arguably the simplest one.  Further properties were commented
in the original paper \cite{BSPME00}.  In particular, note that
(\ref{Pi1}) need not be normalized, and in fact may be not
normalizable, as may also be the case classically, e.g.  because of
periodic crossings in a harmonic potential. It can also be defined even if the system is not classically integrable.

\subsection{Interacting case: second proposal}

The other two definitions discussed here are applicable to 
``scattering potentials'' where M\"oller operators exist, 
\beq\label{opm}
\Opm=\lim_{t\to\mp\infty} e^{i\wh{H}t/\hbar}e^{-i\wh{H}_0 t/\hbar}.
\eeq
We shall assume that the potential operator 
$\wh{V}$ has a local  coordinate representation 
\beq
\la x|\wh{V}|x'\ra=\delta(x-x')V(x),
\eeq
with potential function $V(x)$ vanishing at long distances
from the interaction region so that the (strong) limits in Eq. (\ref{opm})
exist. 
As usual, we shall also consider, with the same notation,
extensions of these Hilbert 
space operators that can be applied to plane waves,
\beq
|p_\pm\ra=\Opm|p\ra\equiv
|p\ra+\lim_{\vep\to 0^+}\frac{1}{E_p\pm i\vep-\wh{H}}\wh{V}|p\ra,\;\; 
E_p=p^2/2m. 
\eeq
The Lippmann-Schwinger states $|p_\pm\ra$ are improper (not square
integrable) eigenstates of $\wh{H}$ with eigenvalue $E_p$.  $\Opm$ are
in general only isometric, i.e., they conserve norm, but in the
absence of bound states they become also unitary.  We shall limit
ourselves to this later case hereafter.  The physical meaning of
$\Opm$ is best understood with the aid of ``asymptotic'' incoming and
outgoing states, $\phi_{in}$ and $\phi_{out}$, to which the actual
state $\psi$ tends in a strong sense in the infinite past and future,
respectively.  These asymptotic states evolve freely, with $\wh{H}_0$.
Whereas $\Op$ provides the scattering state by acting on the incoming
asymptote, $\Om$ does the same job acting on the outgoing one,
\beqa\label{psiin}
\Op |\phi_{in}(t)\ra&=&|\psi(t)\ra,
\\
\label{psiout}
\Om |\phi_{out}(t)\ra&=&|\psi(t)\ra, 
\eeqa
for all $t$. 
 
Without bound states, the resolution of unity can be written 
equivalently as 
\beq
\wh{1}=\intf dp\, |p\ra\la p|=\intf dp\, |p_\pm\ra\la p_\pm|,
\eeq
and the Hamiltonians   
$\wh{H}_0$ and $\wh{H}$ may be related
by the  unitary transformation
\beq
\wh{H}={\Opm} \wh{H}_0 {\Opm}, 
\eeq
with a similar relation holding for functions of $\wh{H}_0$ and $\wh{H}$.
This suggests using the crossing states
\beq
|u^\beta_\pm(X)\ra_2=\Opm |u^\beta(X)\ra_K\,.
\eeq
Note the additional subscript in the
crosssing state, $+$ or $-$, due to the possibility to act with either
one of the two M\"oller operators on $|u^\beta\ra_K$.  Thus, this
procedure generates two different distributions labelled by $+$ or
$-$,
\beq
\Pi_{2,\pm}=\sum_\beta|\la \psi(T)|u_\pm^\beta(X)\ra_2|^2.\label{pi2}
\eeq
These distributions previously appeared in \cite{LJPU99} (later superseded by \cite{LJPU00}). It is important to notice that this is not the way in which the
original distributions in \cite{LJPU99} were presented; they have been adapted
here to the unified notation we are using in order to achieve a better
handle for comparison. 
Their physical content is made evident
by making use of   
(\ref{psiin}), (\ref{psiout}), and the isometry of the M\"oller operators,
\beqa
\la\psi(T)|\Op|u^\beta(X)\ra_K&=&  
\la\phi_{in}(T)|\Op\da\Op|u^\beta(X)\ra_K=
\la\phi_{in}(T)|u^\beta(X)\ra_K,
\\
\la\psi(T)|\Om|u^\beta(X)\ra_K&=&  
\la\phi_{out}(T)|\Om\da\Om|u^\beta(X)\ra_K=
\la\phi_{out}(T)|u^\beta(X)\ra_K.
\eeqa
This means that the generated distributions, $\Pi_{2,+}$ and 
$\Pi_{2,-}$ are nothing but the Kijowski distributions corresponding 
to the incoming and outgoing free-motion asymptotes, respectively.  
These may be useful objects before ($\Pi_{2,+}$) and after 
($\Pi_{2,-}$) the collision, but not in the midst of it.

\subsection{Interacting case: Third proposal}
A different, more complex proposal is based on the
following crossing states, 
\beq
|u^\beta_\pm(X)\ra_3
=\Opm\Theta(\alpha\wh{p})(|\wh{p}|/m)^{1/2}\Opm\da|X\ra,
\eeq
which lead again to two different distributions, 
\beq
\Pi_{3,\pm}(T)=\sum_\beta|\la\psi(T)|u^\beta_\pm(X)\ra_3|^2.
\eeq
These distributions were first introduced and discussed by Le\'on et al. in \cite{LJPU00}, as justified by quantization through quantum canonical transformations of the classical time of arrival (see also \cite{Leon00} for further analysis of this proposal). As previously stated, the rewriting in terms of crossing states is intended to clarify the physical consequences of these distributions.
To analyze their meaning, let us first study the 
amplitudes corresponding to $\Op$,
by inserting $\wh{1}=\Op\da\Op$, and rewriting them as  
\beq
\la \psi(T)|\left[\Op\Theta(\alpha\wh{p})\Op\da\right]
\left[\Op(|\wh{p}|/m)^{1/2}\Op\da\right]|X\ra. 
\eeq
The operators involved have been separated in two 
brackets that can be interpreted physically. The first one is
a projector that selects the part of a wave function 
that {\it had} positive ($\alpha=+$) or negative momentum ($\alpha=-$)
in the infinite past, 
\beq
\wh{F}_+^\beta\equiv \Op\Theta(\alpha\wh{p})\Op\da
=\alpha\int_0^{\alpha\infty} dp\,|p_+\ra\la p_+|; 
\eeq
the second group of operators is
\beq
(2\wh{H}/m)^{1/4}\equiv m^{-{1/2}}\Op|\wh{p}|^{1/2}\Op\da
=\intf dp\,|p_+\ra (2E_p/m)^{1/4} \la p_+|,
\eeq
where the positive root is taken.   
For the case in which the incoming asymptote is restricted 
to positive momenta,  
\beqa
\la\psi(T)|\wh{F}_+^L&=&\la\psi(T)|,
\\
\la\psi(T)|\wh{F}_+^R&=&0,
\eeqa
so that the proposed distribution takes the form 
\beq\label{27}
\Pi_{3,+}(T)=\Pi_3^L(T)
=\la\psi(T)|(2\wh{H}/m)^{1/4}\delta(\wh{x}-X)(2\wh{H}/m)^{1/4}|\psi(T)\ra.
\eeq
The resulting operator is a quantum symmetrization 
of the classical phase space dynamical variable 
\beq\label{lecla}
(2E/m)^{1/2}\delta(x-X),
\eeq
$E$ being the {\it total} energy, to be compared with the classical
variables of the first generalization which  
lead to the positive and minus negative fluxes,
\beqa
\label{pos}
(|p|/m)\Theta(p)\delta(x-X),
\\
\label{neg}
(|p|/m)\Theta(-p)\delta(x-X).
\eeqa
Note that, for an $X$ such that $V(X)=0$, 
the (classical) average of (\ref{lecla}) is equal to the average of the 
sum of (\ref{pos}) and (\ref{neg}), i.e. to the time of 
arrival distribution,
\beq
\la (2E/m)^{1/2}\delta(x-X)\ra_{cl}=J_{cl}^L-J_{cl}^R,\;\;\;
{\rm if}\;  V(X)=0.
\eeq
However, in the interaction region, it is in fact proportional to the
local square root of the total energy.  This may lead to significant
differences with the actual time-of-arrival distribution, which
classically is always given by $J_{cl}^L-J_{cl}^R$ irrespective of the
value of $V(X)$.  In particular, for a potential barrier such that the
initial (asymptotic) momenta of the particles slow down on its top to
a smaller value, (\ref{lecla}) will overestimate the value of the true
arrival time distribution, since for each trajectory the square root
$(2mE)^{1/2}$ is used, instead of the smaller local momentum that
determines the partial fluxes.  Note also that the superscript $L$ in
(\ref{27}) does not make reference here to actual crossing from the
left in the classical limit, but to motion from the left {\it in the
infinite past}. In this respect, (\ref{lecla}) equally counts right or
left crossings at a given time.

A similar analysis may be carried out for a state with incoming
asymptote in the subspace of negative momenta. Thus, for an arbitrary
initial state with positive and negative momentum components,
$\Pi_{3,+}=\Pi_{3,+}^L+\Pi_{3,+}^R$ provides a quantum version of the
distribution that corresponds to the classical dynamical variable
(\ref{lecla}). Repeating the same steps with $\Om$, $\Pi_{3,-}^\beta$
may be interpreted as the quantum versions of the classical
distributions of (\ref{lecla}) for particles that will have positive
or negative momenta in the infinite future.

\section{Numerical example}

We shall illustrate the results of the previous section with 
the collision of a wave packet for a particle of mass $m=1$
with a square barrier.  
The initial state (at $t=0$) is chosen as a  minimum-uncertainty-product 
Gaussian with negligible negative momentum components. Its average position,
average momentum, and standard deviation     
are for the first four figures, in atomic units, 
\beq\label{par1}
\la\wh{x}\ra =-6,\;\; \la \wh{p}\ra =6,\;\; \Delta x=1,
\eeq
whereas the barrier energy, initial and final points are   
\beq\label{par2}
V_0=10,\;\; x_i=0,\;\; x_f=10.  
\eeq
With these parameters the barrier is rather opaque.  This prevents
tunneling and makes the collision predominantly classical.  The
figures combine variously the different distributions, their
components, and the flux for the three points, $X=-2,5,12$,
corresponding to positions before, in, and after the barrier
respectively.  The evaluation of the integrals requires explicit
representations of the states $|p_\pm\ra$ that can be found elsewhere
\cite{BSM94}.

\begin{figure}[h]
\epsfysize=8cm
\centerline{\epsfbox{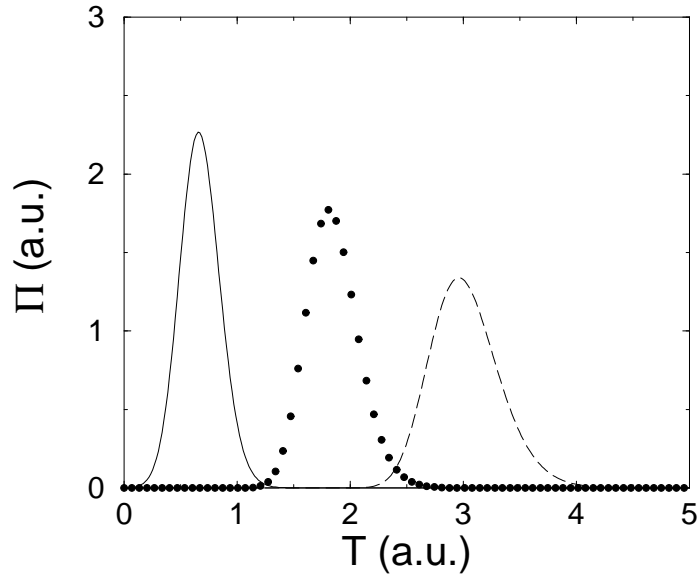}}
\caption{$\Pi_K$ and $\Pi_{2,+}$ (which overlap) for $X=-2,5$
and $12$ (solid, dotted, and dashed lines)
for the Gaussian wave packet described in the text, see (\ref{par1}).}
\end{figure}

Fig. 1 represents Kijowski's distribution for the case in which the
wave packet evolves freely, without interaction with the barrier, as
well as $\Pi_{2,+}$ in the interacting case.  Since the fraction of
negative momenta in the initial state is negligible, it may be
considered essentially equal to the corresponding incoming asymptotic
state for the numerical accuracy of the figures.  Thus $\Pi_{2,+}$ (or
$\Pi_{2,+}^L$) and $\Pi_K$ are indistinguishable for the three values
of $X$.

Fig. 2 shows $\Pi_{2,-}^L$ and the flux $J$, before, in, and after the
barrier, which is now present. They coincide after the barrier, for
$X=12$, but differ otherwise. This is to be expected since
$\Pi_{2,-}^L$ is a free-motion Kijowski distribution for the positive
momentum part of the outgoing asymptotic state.

\begin{figure}[h]
\epsfysize=8cm
\centerline{\epsfbox{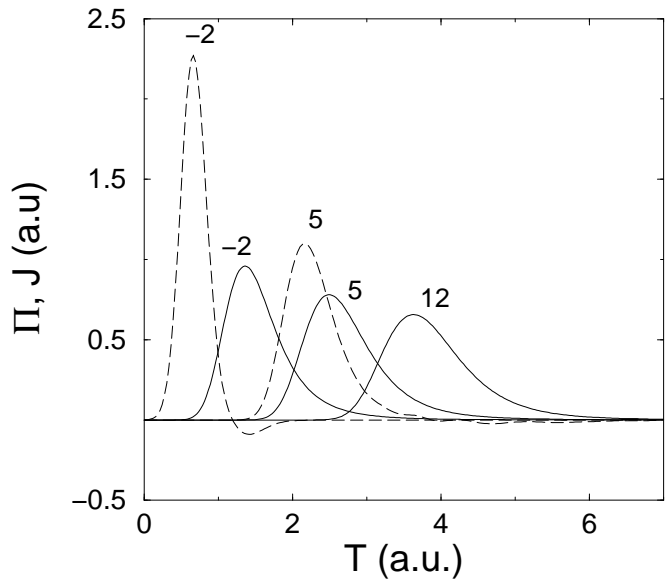}}
\caption{$\Pi_{2,-}^L$ and $J$ (solid and dashed lines respectively)
for $X=-2,5$ and $12$ (indistinguishable in the later case).}
\end{figure}

At $X=-2$, where the potential vanishes, $\Pi_1$ and $\Pi_{3,+}$ are
in essential agreement, see Fig. 3. The second, smaller bump
corresponds to the period of negative flux due to a small reflection.
This smaller bump is due to a contribution of $\Pi_1^R$ only, with no
contribution from $\Pi_1^L$, since this negative flux is associated
with crossing from the right.  However, $\Pi_{3,+}^L$ alone (which in
the figure is indistinguishable from the total $\Pi_{3,+}$) provides
incident and reflected bumps even though they are associated with
different crossing directions. Recall in this respect that the
superscript $L$ in $\Pi_{3,+}^L$ means ``motion from the left (positive
momentum) {\it in the infinite past}''.  (The reflected part will be
seen in more detail in a different collision described in Fig. 5
below.)  The difference $\Pi_1^L-\Pi_1^R$ has also been depicted; note
its agreement with the flux.
    
\begin{figure}[h]
\epsfysize=8cm
\centerline{\epsfbox{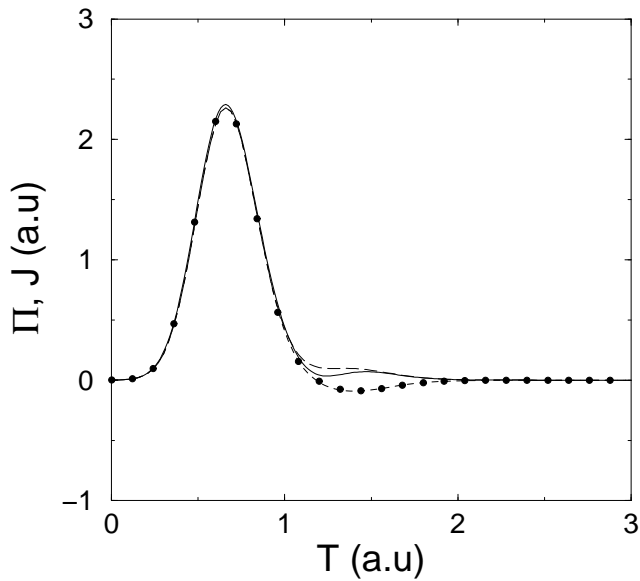}}
\caption{$J$ (dots), $\Pi_{3,+}$ (solid line), $\Pi_1$ (long dashed line), 
and $\Pi_1^L-\Pi_1^R$ (short dashed line)
for $X=-2$.}
\end{figure}

At $X=12$ there is only transmission and $\Pi_1$, $\Pi_{3,+}$ and $J$
are indistinguishable.  However, for $X=5$,
i.e. in the barrier, $\Pi_{3,+}$ is clearly larger than $J$ and
$\Pi_1$, which essentially coincide, see Fig. 4.  A simple estimate of
the ratio between the peaks follows from the classical limit discussed
in the previous section: $\Pi_1$ must be to a good approximation
proportional to an average local momentum; with the current parameters
this entails $[2(18-10)]^{1/2}=4$. It is important to notice that the
relevant quantity is the total energy (taken as $\approx 18$ from the
initial value of the momentum) {\it minus} the potential barrier
energy.  On the other hand, $\Pi_{3,+}$ is proportional to the square root
of the total energy, $(2\times18)^{1/2}=6$.  The ratio of the two
peaks in the figure is indeed 6:4.

\begin{figure}[h]
\epsfysize=8cm
\centerline{\epsfbox{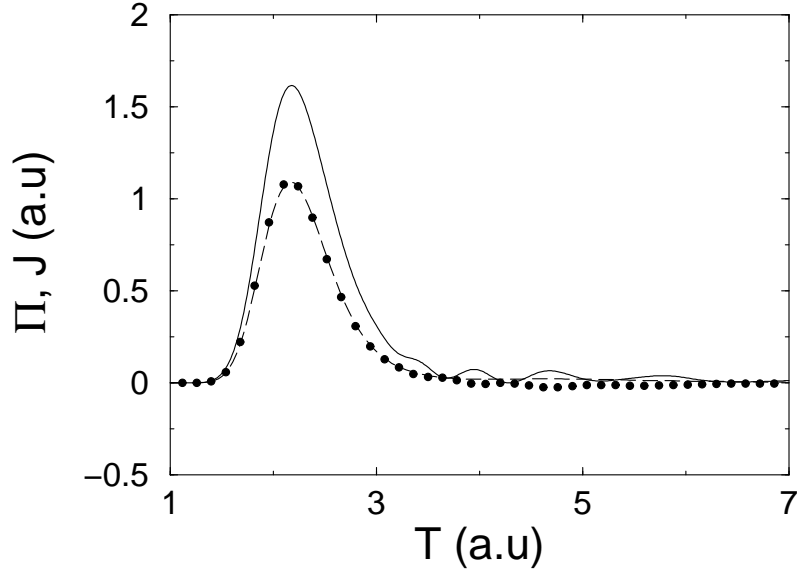}}
\caption{$J$ (dots), $\Pi_{3,+}$ (solid line), and $\Pi_1$ (dashed line) 
for $X=5$.}
\end{figure}

As a complement to Figure 3, we have lowered the average momentum of
the initial wave packet to $\la \wh{p}\ra=3$, while keeping $\Delta
x=1$, so that the whole packet is now reflected.  We evaluate $J$ and
all distributions at $x=-5$ for an initial average position $\la
\wh{x}\ra=-9$. The flux, as portrayed in Fig. 5, shows clearly a positive part 
during the incidence and a negative part corresponding to reflection.
Since several combinantions of the different distributions and their
components match with adequately the two bumps of $|J|$ or just one of
those, we have in fact only represented $\Pi_{3,+}=\Pi_{3,+}^L$, with
a dashed line.  $\Pi_{1}$ is barely distinguishable from it. The
incidence bump on the left is also reproduced by $\Pi_{2,+}$, or by
$\Pi_1^L$, whereas the reflection bump is reproduced by $\Pi_{2,-}$ or
$\Pi_1^R$.

\begin{figure}[h]
\epsfysize=8cm
\centerline{\epsfbox{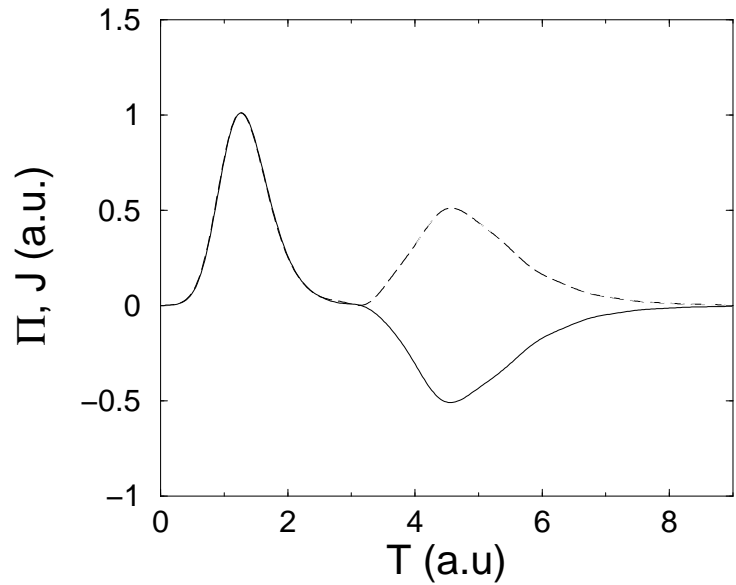}}
\caption{$J$ (solid line) and $\Pi_{3,+}$
(dashed line) for $X=-5$. See the
text for details.}
\end{figure}

\section{Conclusion}
We have examined three different generalizations for Kijowski's
time-of-arrival distribution in the interacting case, both formally
and numerically. This exam has yielded the result that, among these
three, there is only one, $\Pi_1$, in fact the simplest, that
satisfies the correspondence principle in the sense of recovering the
classical expression for the time of arrival when the effects of
non-commutativity of the operators involved may be neglected.  The
other two proposals provide the correct classical limit in certain
cases, but not in general.  The numerical analysis of these
distributions supports these formal considerations.

In order to make an adequate comparison between these three proposals, 
it has proved convenient to write them in the unified formalism of the 
``crossing states''.  The formalism itself suggests further 
generalizations, by considering alternative crossing states.  An open 
question is whether the crossing state formalism is indeed the most 
adequate one for the description of times of arrival, or whether other 
presentations are more suitable.  What is assured is both that the crossing state formalism guarantees covariance and positivity, and its power for comparing widely diverging previous proposals, this being a property that  we would expect to hold more generally.
\acknowledgments

We are grateful to Juan Le\'on, Jaime Julve, and Fernando de Urr\'{\i}es for comments and for checking independently the numerical results for $\Pi_{3,+}$.
This work has been supported
by Ministerio de Educaci\'on
y Cultura (Grants
PB97-1482 and AEN99-0315), The
University of the Basque Country (Grant UPV 063.310-EB187/98), 
and the Basque Government (Grant PI-1999-28).
A. D. Baute acknowledges an FPI fellowship by Ministerio de
Educaci\'on y Cultura.

\end{document}